# Coupled Bloch-Wave Analysis of Active PhC Waveguides and Cavities

Marco Saldutti,[1] Jesper Mørk[2], Paolo Bardella[1], Ivo Montrosset[1] and Mariangela Gioannini[1]

[1] *Dipartimento di Elettronica e Telecomunicazioni, Politecnico di Torino, Corso Duca degli Abruzzi 24, Torino, Italy*
[2] *DTU Fotonik, Department of Photonics Engineering, Technical University of Denmark, Ørsteds Plads, Building 343, DK-2800 Lyngby, Denmark*
marco.saldutti@polito.it

*Abstract—* A coupled Bloch-wave approach is employed to analyze active photonic-crystal (PhC) waveguides and cavities. Gain couples the otherwise independent counter-propagating Bloch modes. This coupling is shown to limit the maximum attainable slow-light enhancement of gain itself and to strongly affect the mode selection in PhC lasers.

## Introduction

The slow-light (SL) enhancement of gain in photonic-crystal waveguides allows for the fabrication of shorter devices when realizing active structures. In particular, PhC lasers based on line-defect waveguides are ideal candidates for energy efficient light sources in high density PhC integrated circuits [1,2]. Solving Maxwell equations by a finite-difference-time-domain (FDTD) technique is a rigorous, but extremely time- and memory-consuming approach to analyze PhC devices [3]. Furthermore, FDTD simulations are not always useful to shed light on the physics of the investigated structures. Conversely, coupled-mode theory has been widely used to investigate the impact of SL effects in both passive [4,5] and active [6] PhC waveguides. In particular, the complex optical susceptibility arising by the interaction of the field with the active medium is treated in [6] as a weak perturbation of the passive structure, which induces a coupling between the otherwise independent counter-propagating Bloch modes. The fundamental limitations to the SL gain-enhancement imposed by the gain itself have been investigated in [7] by a rigorous, non-perturbative approach. In this work we use the perturbative approach of [6] to study an active PhC waveguide; we analyze the implications of the gain perturbation on the group index and then we study a PhC laser modelled as a cavity consisting of an active PhC waveguide and two mirrors. Interestingly, it is shown that our model predicts, consistently with [7], a reduction of the maximal group index caused by increasing the gain and it can be used to understand the impact of the gain-induced coupling on the selection of PhC laser lasing mode.

## I. Numerical Model

The forward- (+) and backward-propagating (−) guided electric field of the passive waveguide in the frequency-domain are denoted by $E_{0,\pm}(r,\omega) = e_{0,\pm}(r,\omega)e^{\pm ik_z(\omega)z}$, where $z$ is the propagation direction and $e_{0,\pm}(x,y,z) = e_{0,\pm}(x,y,z+a)$ are the Bloch waves, with $k_z$ propagation constant and $a$ the PhC lattice constant. The electric field of the active waveguide is expanded as $E = \psi_+(z,\omega)E_{0,+} + \psi_-(z,\omega)E_{0,-}$, where $\psi_\pm(z,\omega)$ are slowly-varying amplitudes. By neglecting nonlinear effects, two coupled differential equations for $\psi_\pm(z,\omega)$ are derived [6]:

$$\begin{cases} \frac{\partial \psi_+(z,\omega)}{\partial z} = i\kappa_{11}(z,\omega)\psi_+(z,\omega) + i\kappa_{12}(z,\omega)e^{-i2k_z(\omega)z}\psi_-(z,\omega) \\ -\frac{\partial \psi_-(z,\omega)}{\partial z} = i\kappa_{21}(z,\omega)e^{i2k_z(\omega)z}\psi_+(z,\omega) + i\kappa_{11}(z,\omega)\psi_-(z,\omega) \end{cases} \quad (1)$$

The self- and cross-coupling coefficients induced by the active material gain $g_0(\omega)$ are indicated as $\kappa_{11;12;21}(z,\omega) \simeq -\frac{i}{2}g_0(\omega)[n_g(\omega)/n_s]\Gamma_{xy,11;12;21}(z,\omega)$, where $n_s$ and $n_g$ are the slab material refractive index and the passive waveguide group index. Confinement factors $\Gamma_{xy,11;12;21}(z,\omega)$ are given by

$$\Gamma_{xy,11}(z,\omega) = \frac{a\int_S \epsilon_0 n_s^2 |e_0(r,\omega)|^2 F(r)dS}{\int_V \epsilon_0 n_b^2(r)|e_0(r,\omega)|^2 dV}$$

$$\Gamma_{xy,12}(z,\omega) = \frac{a\int_S \epsilon_0 n_s^2 [e_{0,-}(r,\omega)\cdot e_{0,+}^*(r,\omega)]F(r)dS}{\int_V \epsilon_0 n_b^2(r)|e_0(r,\omega)|^2 dV}$$

with $\Gamma_{xy,21} = \Gamma_{xy,12}^*$; $V$ is the volume of a PhC supercell, $S$ the transverse plane at position $z$ and $n_b(r)$ the background refractive index, whereas $F(r) = 1 (= 0)$ in the slab (holes). Due to the $z$-periodicity of $e_{0,\pm}$ and $F(r)$, the coupling coefficients are periodic with $z$. If the single unit cell is discretized with a sufficiently small space step $\Delta_z$, the coupling coefficients can be assumed constant within it. By defining $c_\pm = \psi_\pm e^{\pm ik_z z}$, Eq. (1) is turned, in each $\Delta_z$, into an initial-value problem, whose solution in matrix form is

$$\begin{bmatrix} c_+(z_0+\Delta_z) \\ c_-(z_0+\Delta_z) \end{bmatrix} = \begin{bmatrix} T_{\Delta_z,11} & T_{\Delta_z,12} \\ T_{\Delta_z,21} & T_{\Delta_z,22} \end{bmatrix} \begin{bmatrix} c_+(z_0) \\ c_-(z_0) \end{bmatrix} \quad (2)$$

with

$T_{\Delta_z,11;22} = \cosh[\gamma(z_0)\Delta_z] \pm i\frac{\kappa_{11}(z_0)+k_z}{\gamma(z_0)}\sinh[\gamma(z_0)\Delta_z]$,

$T_{\Delta_z,12;21} = \pm i\frac{\kappa_{12;21}(z_0)}{\gamma(z_0)}\sinh[\gamma(z_0)\Delta_z]$, and

$\gamma(z_0) = \sqrt{\kappa_{12}(z_0)\kappa_{21}(z_0) - [\kappa_{11}(z_0)+k_z]^2}$.

By successive application of Eq. (2), the single unit cell transmission matrix $T_a$ is obtained and the transmission matrix of $N$ cascaded cells is given by $T_a^N$. From Frobenius theorem, $T_a^N$ can be written as $T_a^N = M\lambda^N M^{-1}$, where $M$ contains the eigenvectors of $T_a$ arranged by columns and $\lambda$ is a diagonal matrix with the eigenvalues of $T_a$ on the main diagonal. Multiplying by $M^{-1}$ both sides of

$$\begin{bmatrix} c_+(Na) \\ c_-(Na) \end{bmatrix} = M\lambda^N M^{-1} \begin{bmatrix} c_+(0) \\ c_-(0) \end{bmatrix} \quad (3)$$

the Bloch waves of the active waveguide at input and output are obtained, i.e. $c_B(Na) = \lambda^N c_B(0)$. From here, it is apparent that $\lambda^N$ is the evolution matrix in the basis of the Bloch waves of the active waveguide. If the eigenvalues of $T_a$ are denoted by $\lambda_{1,2} = e^{\pm i\phi}$, $\phi = \cos^{-1}[Tr(T_a)/2]$ is the dispersion relation of the active waveguide and $n_{g,Pert}(\omega,g_0) = c\,Re\{\partial\phi(\omega,g_0)/\partial\omega\}$ the associated group index, with $c$ vacuum light speed. Within this approach, a PhC laser consists in the cascade of an active PhC waveguide and two mirrors, which, for simplicity, are modelled as standard reflectors. The complex round-trip-gain (RTG) of the cavity is computed as the product, at a given reference plane, of the left and right field reflectivity. Longitudinal resonant modes are those for which ∠RTG is an integer multiple of $2\pi$. For each longitudinal mode, threshold gain is found as the smallest $g_0$ value which ensures |RTG| = 1 [8].

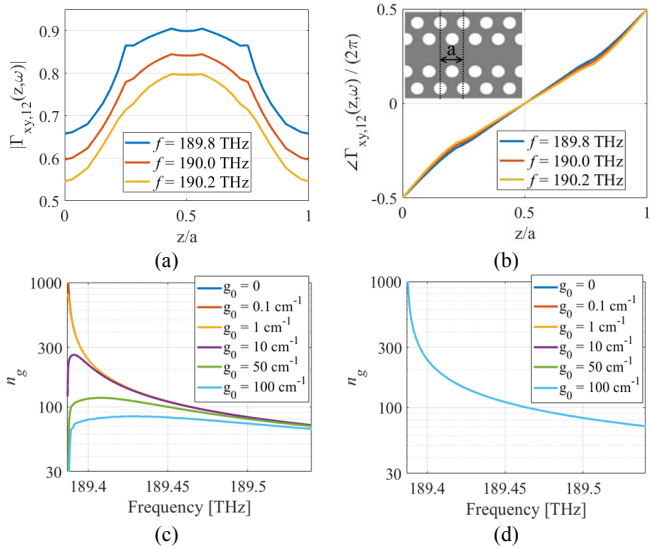
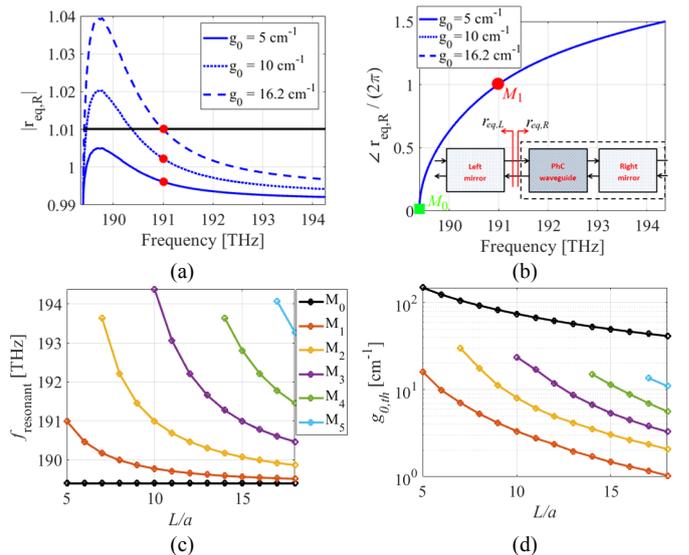

Fig. 1. (a) Magnitude and (b) phase of $\Gamma_{xy,12}$, at different frequencies, for the same line-waveguide of PhC lasers in [2]; inset in (b): unit cell reference planes. Group index with (c) and without (d) gain-induced coupling.

Fig. 2. Magnitude of $r_{eq,R}$, at different $g_0$, for $L = 5a$; black line is level $1/r$. (b) Phase of $r_{eq,R}$; inset: scheme of principle of the cavity (c) Mode frequencies versus cavity length. $M_0$ is the mode at the band-edge. (d) Threshold gain for the onset of lasing of the various modes.

## II. SIMULATION RESULTS

The reference structure is the line-defect waveguide on which the PhC lasers realized in [2] are based. Dispersion relation and Bloch modes of the passive waveguide are computed by the free software package MIT Photonic-Bands (MPB) [9]. Fig. 1a and 1b display magnitude and phase of $\Gamma_{xy,12}(z,\omega)$ at different frequencies. Since the $z$-variation of $\angle\Gamma_{xy,12}(z,\omega)$ on a unit cell is approximately linear with a slope equal to $2\pi/a$, the first-order Fourier component of $\kappa_{12}(z,\omega)$, which synchronously couples $\boldsymbol{E}_{0,+}$ and $\boldsymbol{E}_{0,-}$, is proportional to $g_0(\omega)\,[n_g(\omega)/n_s] <|\Gamma_{xy,12}(z,\omega)|>$. Since $<\Gamma_{xy,11}(z,\omega)>$ and $<|\Gamma_{xy,12}(z,\omega)|>$ have comparable values, the magnitude of the cross-coupling coefficients is comparable with the self-coupling coefficient. This peculiar characteristic of the active PhC waveguides arises from the $2\pi$ phase shift of the non-negligible $z$-component of $\mathbf{e}_{0,\pm}$ along the propagation direction. Fig. 1c reports the group index $n_{g,Pert}$ of the active waveguide as a function of frequency at different $g_0$ values. At small gain values, the dispersion relation of the active waveguide is not significantly perturbed, and the group index diverges as the frequency approaches the band-edge of the passive waveguide, i.e. $k_z a/2\pi = 0.5$ with a frequency $f \approx 189.387$ THz. On the contrary, at larger gain values the group index is reduced and it even starts to decrease in the close proximity of the band-edge. Remarkably, this behaviour is consistent with that reported in [7] and obtained with a non-perturbative treatment. Furthermore, if the gain-induced coupling is neglected (i.e., $\kappa_{12;21} = 0$), the group index monotonically diverges with the frequency approaching the band-edge (Fig.1d). This proves the key role played by cross-coupling in limiting the maximum attainable SL enhancement of gain. With this coupled Bloch-wave approach we have then modelled the PhC lasers presented in [2]. The mirrors reflectivity is set to $r^2 = 0.98$ [10] and $g_0$ is assumed to be frequency-independent. The inset of Fig. 2b displays a scheme of principle of the cavity, with the field reflectivity from the left facet towards the cavity denoted by $r_{eq,R}$. Fig. 2a and 2b focus on a cavity with length $L = 5a$, showing magnitude and phase of $r_{eq,R}$ versus frequency at increasing $g_0$ values. The threshold condition corresponds to the level $1/r$, corresponding to the horizontal line in Fig. 2a. The red spots track the longitudinal resonant mode $M_1$, with frequency $f = 191$ THz, as it approaches the lasing onset at $g_0 = 16.2$ cm$^{-1}$. The mode located exactly at the band-edge ($M_0$, shown in Fig. 2b) requires higher gain for achieving threshold, because the maximum attainable $|r_{eq,R}|$ around the band-edge is limited by the gain induced cross-coupling. This is a consequence of the fact that the field backscattered by the waveguide and the field backscattered by the right mirror facet are out of phase at the band-edge. Fig. 2c,d report, at each cavity length, all the longitudinal modes and corresponding threshold gain. The frequency shift of mode $M_1$ towards the SL region observed by increasing cavity length (Fig. 2c) well reproduces the experimental [2] and numerical [3] trends. Conversely, without the gain-induced distributed feedback, the group index and the effective gain resulting from SL enhancement would monotonically increase towards the band-edge; consequently, the cavity would behave as a SL enhanced FP laser and, independently of the cavity length, the mode M$_0$ would be the sole lasing one.

## III. CONCLUSIONS

In conclusion, the gain-induced coupling between counter-propagating Bloch modes has been found to be responsible for the degradation of the SL enhancement of gain discussed in [7]. Moreover, this coupling strongly affects the lasing mode threshold gain properties of PhC lasers.